\documentclass[aps,prd,10pt,twocolumn,nofootinbib,groupedaddress,amsfonts,floatfix]{revtex4-1}

\usepackage[colorlinks=true,urlcolor=blue,linkcolor=,citecolor=blue]{hyperref}
\usepackage{amsmath,amssymb,amstext,amsbsy,amsfonts,amsthm,graphicx,microtype,dsfont}
\usepackage[capitalize]{cleveref}

\newcommand{\abs}[1]{{\left \vert #1 \right \vert}}

\usepackage{color}
\usepackage{ifthen}
\newboolean{editorial}
\setboolean{editorial}{true}
\newcommand{\editorial}[2]{\ifthenelse{\boolean{editorial}}{\textcolor{red}{[\textsf{\textbf{{#1}}}: }\textcolor{blue}{\textsf{{#2}}}\textcolor{red}{]}}{}}

\usepackage{xcolor}

\newcommand{\mpl}{m_\mathrm{pl}}

\begin{document}

\title{Gravitational waves from asymmetric oscillon dynamics?}

\author{Mustafa~A.~Amin${}^{1}$}
\author{Jonathan~Braden${}^{2}$}
\author{Edmund~J.~Copeland${}^{3}$}
\author{John~T.~Giblin,~Jr${}^{4,5}$}
\author{Christian~Solorio${}^{4}$}
\author{Zachary~J.~Weiner${}^{6}$}
\author{Shuang-Yong~Zhou${}^{7}$}
\affiliation{${}^1$Department of Physics \& Astronomy, Rice University, Houston, Texas 77005-1827, U.S.A.}
\affiliation{${}^2$Department of Physics \& Astronomy, University College London, Gower Street, London, WC1E 6BT, UK}
\affiliation{${}^3$School of Physics and Astronomy, University of Nottingham Nottingham, NG7 2RD, United Kingdom}
\affiliation{${}^4$Department of Physics, Kenyon College, Gambier, OH 43022, U.S.A.}
\affiliation{${}^5$CERCA/ISO, Department of Physics, Case Western Reserve University, Cleveland, OH 44106, U.S.A.}
\affiliation{${}^6$Department of Physics, University of Illinois at Urbana-Champaign, Urbana, IL 61801, U.S.A.}
\affiliation{${}^7$Interdisciplinary Center for Theoretical Study, University of Science and Technology of China, Hefei, Anhui 230026, China}

%%%%%%%%%%%%%%%%%%%%%%%%%%%%%%%%%%%%%%%%%%%%%%%%%%%%%%%%%%%

\begin{abstract}

It has been recently suggested that oscillons produced in the early universe from certain {\it asymmetric} potentials continue to emit gravitational waves for a number of $e$-folds of expansion {\it after} their formation, leading to potentially detectable gravitational wave signals.
We revisit this claim by conducting a convergence study using graphics processing unit (GPU)-accelerated lattice simulations and show that numerical errors accumulated with time are significant in low-resolution scenarios, or in scenarios where the run-time causes the resolution to drop below the relevant scales in the problem.
Our study determines that the dominant, growing high frequency peak of the gravitational wave signals in the fiducial ``hill-top model" in~\cite{Antusch:2016con} is a numerical artifact.
This finding prompts the need for a more careful analysis of the numerical validity of other similar results related to gravitational waves from oscillon dynamics.
\end{abstract}

\maketitle

%%%%%%%%%%%%%%%    Introduction     %%%%%%%%%%%%%%%%%%%%%%%%%%%%%%

\section{Introduction}

Gravitational wave production in the post-inflationary universe is typically sourced by nonperturbative dynamics of cosmological fields. The violent dynamics of energy transfer at the end of inflation, as well as phase transitions in the early universe are commonly explored sources of a primordial, stochastic background of gravitational waves (\cite{Khlebnikov:1997di,Easther:2006gt,Easther:2006vd,Dufaux:2007pt, Hindmarsh:2013xza, Caprini:2015zlo}; for a recent review, see~\cite{Caprini:2018mtu}). Amongst postinflationary sources, the formation and dynamics of solitonic configurations called {\it oscillons} can also provide an exciting high-frequency source of stochastic gravitational waves~\cite{Zhou:2013tsa, Antusch:2016con, Antusch:2017flz, Antusch:2017vga, Liu:2017hua}.

Oscillons are localized, mostly spherical, nontopological quasisolitons~\cite{Bogolyubsky:1976yu,Gleiser:1993pt,Copeland:1995fq,Amin:2010jq}, which are very long lived. Heuristically speaking,  for oscillons to exist, the field potential must have a quadratic minimum, and be shallower than quadratic as one moves away from the minimum (for a more precise and general condition, see for example~\cite{Amin:2013ika}).
Recent observations~\cite{Mortonson:2010er, Ade:2015lrj} have confirmed the possibility of having shallower than quadratic potentials during inflation, which is fascinating because such oscillon-producing inflationary potentials arise in a number of well-motivated models~\cite{Silverstein:2008sg, McAllister:2008hb, Dong:2010in, Kallosh:2010xz}. If the field-distance where the potential becomes shallower than quadratic is small compared $m_{\rm pl}$, it is possible that at the end of inflation there was copious production of oscillons from the inflaton field (for example, see~\cite{Amin:2011hj,Gleiser:2014ipa}).

Gravitational wave production from the post-inflationary formation and evolution of oscillons has recently garnered attention in the literature~\cite{Zhou:2013tsa, Antusch:2016con, Antusch:2017flz, Antusch:2017vga, Liu:2017hua}. In~\cite{Antusch:2016con}, the authors claim signals loud enough to intersect the observational capabilities of the Laser Interferometer Gravitational-Wave Observatory (LIGO)~\cite{Abbott:2017xzg,Sato:2017dkf}. This would of course be an extremely exciting result, a direct probe of the earliest moments of the Universe. Here we revisit this intriguing claim for the fiducial ``hill-top model'' used in~\cite{Antusch:2016con}. By using a convergence study examining the characteristics of the predicted signal as we vary numerical parameters, we provide a robust test regarding the existence or nonexistence of a detectable signal, and with it demonstrate that the signal claimed in~\cite{Antusch:2016con} has to be a numerical artifact. Perhaps of greater significance, we show that the tests we are proposing are crucial when analyzing compact structures for long run times, hence have applications beyond oscillons to any compact nonlinear features which can have long lifetimes in an expanding universe. 

In what follows we begin by describing the inflationary model used in~\cite{Antusch:2016con}, followed by a description of the scheme by which we calculate the gravitational wave signal. We then describe our numerical methods and show results for a set of resolutions, constituting the convergence study.

%%%%%%%%%%%%%%%    The Model     %%%%%%%%%%%%%%%%%%%%%%%%%%%%%%

\section{The model}

We consider an inflation model where a canonical scalar field is minimally coupled to Einstein gravity
\begin{align}
	\mathcal{L} = \sqrt{-g} \left[ \frac{m_\text{pl}^2}{2} R - \frac{1}{2} \partial_\mu \phi \partial^\mu \phi - V(\phi) \right] ,
    \label{continuumaction}
\end{align}
with the potential
\begin{align}
	V(\phi) = V_0 \left( 1 - \frac{\phi^p}{v^p} \right)^2 ,
	\label{infpot}
\end{align}
where $m_\text{pl}$ is the reduced Planck mass, $R$ is the Ricci scalar, and $\phi$ is a real scalar field.
To compare to the results of~\cite{Antusch:2016con}, we set $p=6$, $v=10^{-2} \mpl$, and $V_0 = 10^{-13}v^3 \mpl$. These were chosen to be roughly consistent with observations of the amplitude and spectral index of the observed CMB anisotropies, assuming $\phi$ is the inflaton. The effective mass of the field at the minimum of the potential $m_\phi = p \sqrt{2V_0/v^2} \approx 2.68 \times 10^{-7} m_{\rm pl}$.

We work in the flat Friedmann-Lema\^itre-Robertson-Walker (FLRW) background where gravitational waves are defined as the transverse and traceless perturbation $h_{ij}$ of the metric
\begin{align}
	ds^2 = - dt^2 + a(t)^2 \left( \delta_{ij} + h_{ij} \right) dx^i dx^j ,
\end{align}
where $h_{ii} = \partial_i h_{ij} = 0$. Latin indices are contracted with $\delta_{ij}$.
The scale factor $a(t)$ evolves according to Friedmann's equation,
\begin{align}\label{friedmann}
	H(t)^2
	&\equiv \left(\frac{\dot{a}}{a}\right)^2
	= \frac{ 1 }{3 m_\text{pl}^2} \rho(t),
\end{align}
where $\rho$ is the average energy density---that is, the 00 component of the stress-energy tensor $T_{00}$ averaged over the spatial volume,
\begin{align}
	\rho \equiv \left\langle T_{00} \right\rangle
	&= \left\langle \frac{1}{2} {\dot{\phi}}^2 + \frac{1}{2 a^2} (\nabla\phi)^2 + V(\phi)  \right\rangle.
\end{align}
Here overdots denote derivatives with respect to the cosmic time $t$ and $\nabla$ is the spatial gradient.
The scalar field satisfies the Klein-Gordon equation
\begin{align}\label{phiEOM}
	\ddot{\phi} = \frac{\nabla^2 \phi}{a^2} - 3 H \dot{\phi} - \frac{dV}{d\phi}.
\end{align}

%%%%%%%%%%%%%%%%%%%%%%%%%%%%%%%%%%%%%%%%%%%%%%%%%%%%%%%%%%%

\section{Generation of gravitational waves}

The linearized Einstein equation for gravitational waves is given by
\begin{align}\label{hEOM}
	\ddot{h}_{ij} = \frac{ \nabla^2 h_{ij} }{a^2} - 3 H \dot{h}_{ij} + \frac{2}{m_\text{pl}^2 a^2} S_{ij}.
\end{align}
where $S_{ij}$ is the transverse-traceless projection of the energy momentum tensor
\begin{align}
	S_{ij} = \left( P_{ik} P_{jl} - \frac{1}{2} P_{ij} P_{kl} \right) T_{kl} ,
\end{align}
with
\begin{align}\label{projector}
	P_{ij} = \delta_{ij} - \frac{k_i k_j}{\abs{k}^2}.
\end{align}
The stress tensor of the scalar field is
\begin{align}
	T_{ij} = \partial_i \phi \partial_j \phi - g_{ij} \mathcal{L}.
\end{align}
Because $g_{ij} \mathcal{L}$ is proportional to $g_{ij}$, it is part of the trace of the stress tensor; for this reason it suffices to compute $\partial_i \phi \partial_j \phi$ prior to projecting to obtain $S_{ij}$.

The effective stress-energy tensor for gravitational waves is given by~\cite{Misner:1974qy}
\begin{align}
	T_{\mu\nu}^\text{gw} = \frac{1}{4} m_\text{pl}^2 \left\langle \partial_\mu h_{ij} \partial_\nu h_{ij} \right\rangle,
\end{align}
the 00 component of which gives the energy density
\begin{align}
	\rho_\text{gw} = \sum_{i,j} \frac{m_\text{pl}^2}{4} \left\langle \dot{h}_{ij}^2 \right\rangle.
\end{align}

By Parseval's theorem and differentiation, we get
\begin{align}
	\frac{d\rho_\text{gw}}{d\ln k} = \sum_{i,j} \frac{\pi m_\text{pl}^2}{L^3} k^3 \abs{ \dot{h}_{ij}(k) }^2.
\end{align}
Converting to the fractional energy density $\Omega_\text{gw}$, we have
\begin{align}
	\Omega_\text{gw}(a_{\rm e})
	&\equiv \frac{1}{\rho_\text{crit}} \frac{d\rho_\text{gw}}{d\ln k} \\
	&= \frac{\pi k^3}{3 H^2 L^3} \sum_{i,j} \abs{ \dot{h}_{ij}(k) }^2.
\end{align}
where $L$ is the length of the side of the box. Finally by redshifting the radiation to today, we obtain
\begin{align}\label{transferamp}
	\Omega_\text{gw,0} h_0^2
	&=\Omega_{\rm gw}(a_{\rm e}) \times \left(\frac{g_{\rm th}}{g_0}\right)^{-1/3}\Omega_{{\rm r},0}h_0^2\,,
\end{align}
where $a_{\rm e}$ is the scale factor at which we are calculating the gravitational wave signal, $g_0 / g_{\rm th} = 1/100$ is the ratio of the number of degrees of freedom today to the earlier epoch, $\Omega_{\text{r},0}$ is the present radiation energy density and $h_0$ parametrizes uncertainty in the Hubble parameter today. Note that one always pays a price in terms of the $\Omega_{{\rm gw},0}$ amplitude when the equation of state $w<1/3$ after or during the production of gravitational waves, since \cref{transferamp} assumes that the universe is radiation dominated from $a_\mathrm{e}$ until matter-radiation equality. For the late universe $\Omega_{\mathrm{r},0}$ accounts for this loss of amplitude. For the sake of simplicity we have assumed radiation domination from the time of production to matter radiation equality.
Note that the observed frequency corresponding to a wave vector $k$ is~\cite{Easther:2006gt}
\begin{align}
	f \approx 2.7 \times 10^{10} \frac{k}{a_{\rm e}\sqrt{m_\text{pl} H_{\rm e}}} \text{ Hz}.
\end{align}

%%%%%%%%%%%%%%%%%%%%%%%%%    Numerical Simulations %%%%%%%%%%%%%%%%%%%%%%%%

\section{Numerical simulations}

For this work we use a GPU-accelerated code whose structure uses the logic from {\sc Grid and Bubble Evolver} ({\sc GABE})~\cite{Child:2013ria,gabe} and {\sc LatticeEasy}~\cite{Felder:2000hq}.
We evolve the metric perturbations (which do not backreact onto the scalar field) and output gravitational wave spectra using the same strategy as in~\cite{Easther:2006vd,Easther:2007vj}.
At its heart, this software evolves \cref{phiEOM} on a finite, $N \times N \times N$ grid.
The expansion of the Universe is calculated self-consistently using \cref{friedmann}.
At each step we calculate the anisotropic stress tensor in configuration space, transform it into momentum space, project it onto the transverse-traceless space using the ``continuum'' projector, \cref{projector}, and calculate the source term of \cref{hEOM}.
The metric perturbations themselves are always computed in momentum space.

This numerical procedure differs from the methods of~\cite{Child:2013ria,Easther:2006vd,Easther:2007vj} in three significant ways.
First, in a way similar to~\cite{Easther:2010qz} spatial derivatives are calculated using a pseudospectral method: at every step, the field(s) are Fourier transformed, multiplied by the appropriate component of $i \vec{k}$ (or $- \vert \vec{k} \vert^2$ for Laplacians), and then inverse transformed back to position space.
This calculates (and stores) all numerical derivatives at all points simultaneously and, for our periodic and regularly spaced grid, is the best possible approximation of the continuum derivatives, being exponentially convergent.
We can then use these derivatives for both evolving the fields and calculating the anisotropic stress tensor.
Computing derivatives in this way additionally avoids the cancellation error unavoidable for stencil derivatives, i.e., the loss of precision resulting from subtracting two nearly-equal floating-point numbers.

Second, this code uses a fourth-order Runge-Kutta integration scheme as opposed to the second-order methods employed by \textsc{GABE} and \textsc{LatticeEasy}.
Finally, the code is written in CUDA~\cite{CUDA} to run on a GPU, allowing us to perform traditionally long simulations in much shorter times (by one to two orders of magnitude).
Also, by relying on the \textsc{CUFFT} fast Fourier transform library~\cite{CUDA} for the majority of the computation, the pseudospectral method is more efficient on this hardware (given that it requires more computations per step than a finite-difference method).

pseudospectral derivatives minimize errors generally found for stencil derivatives arising from inconsistency with the use of the continuum-based projector, \cref{projector}---a known issue for stencil derivatives~\cite{Huang:2011gf}.
In this way, our study isolates errors due to lattice resolution from those due to the choice of projector.

To complete our convergence study we focus on a single physical scenario, originally investigated in~\cite{Antusch:2016con}. The authors there chose initial homogeneous field values, $\phi_i^{\rm [1]} = 0.08 v$ and $\dot{\phi}_i^{\rm [1]} = 2.49\times10^{-9}v^2$, that are consistent with the physical values of the field at the end of inflation.
They also chose an initial box with sides, $L_i^{\rm [1]} = 0.01/{H_i^{\rm [1]}}$, where $H_i^{\rm [1]}$ is the initial Hubble parameter in their simulation.
At this point, even though inflation is over, the potential is still concave down. However, the wavelength of the tachyonic modes is much longer than the size of the simulation box at the initial time ($\lambda_{\rm tach}/L_i^{\rm [1]}\gtrsim 70$).

To avoid unnecessary loss of resolution and computational resources, we begin our simulations slightly later than~\cite{Antusch:2016con}. For our study, we have chosen to enforce a stricter condition, requiring that $V''(\phi) \geq 0$ at the initial time (other codes, e.g. {\sc LatticeEasy}, set this effective mass to zero if the modes appear tachyonic).
We have checked that there is no significant time evolution of modes inside the simulation volume between the initial time used in~\cite{Antusch:2016con} and our initial time.

Our initial conditions have homogenous field values $\phi_i \approx 0.877 v$ and $\dot{\phi}_i \approx 3.73 \times 10^{-6} \, v^2$.
We set our initial box size to $L_i = 1.29 \times 0.01 / H_i$, where $H_i$ is the initial Hubble parameter in the current simulations and the factor of $1.29$, obtained by solving the ODE's for the homogeneous mode, sets our physical box size to be the same as in~\cite{Antusch:2016con}.
This means our simulations start (i.e., $a = 1$) at a time when $a^{\rm [1]} = 1.29$ in the simulations of~\cite{Antusch:2016con}, and therefore scalefactors should be translated between the two results accordingly.

For future convenience, we note that $H_i \approx 6.78 \times 10^{-4} m_\phi$ and $L_i \approx 19.05 m_\phi^{-1}$ where $m_\phi^2 = V''(\phi=v)$.
This mass $m_\phi$ sets the essential length/time scale in the following dynamics.
We generate initial fluctuations of the field and its derivative in momentum space following \textsc{GABE} and \textsc{LatticeEasy}'s standard procedure, sampling mode amplitudes from a Rayleigh distribution and phases uniformly on the interval $[0,2\pi)$, so that the field and its derivatives are solutions to the Kline-Gordon equation which reproduce the Bunch-Davies vacuum,
\begin{align}
	\left\langle \phi(\vec{k}) \phi(\vec{k}')^\ast \right\rangle = \frac{1}{2 \omega(\vec{k})} \delta^{(3)}(\vec{k}-\vec{k}'),
\end{align}
where $\omega^2(\vec{k}) = \vert \vec{k} \vert^2 + m_{\rm eff}^2$ with $m_{\rm eff}^2 = V''(\phi=\phi_i)$.

Our convergence study will compare three different resolutions, $N=64$, $N=128$, and $N=256$, to look for physical signals whose characteristics will be independent of numerical parameters in the regions of validity of the simulations, as well as signs of numerical artifacts that creep in as resolution-dependent effects.
Note that our initial lattice spacing is $dx_{\scriptscriptstyle N}= L_i/N$.
All simulations use a time step $dt = dx / 20$.
As the universe expands, the physical value of this spacing becomes $a(t)dx_{\scriptscriptstyle N}$.
We end all of our simulations at $a = 16$, which occurs after $a(t) dx_{\scriptscriptstyle 256} \approx m_{\phi}^{-1}$, i.e., after $a \approx 13.45$ ($2.6$ $e$-folds) or equivalently $a^{\rm [1]}\approx 13.4 \times 1.29$.

For these simulations we use conformal time, $d\tau = a(t) dt$, and are careful not to run the simulation past where we have sufficient time resolution.
In the simulations presented here, the {\sl physical} time step is still smaller than $m_\phi^{-1}$ in all of the simulations.

To provide a direct comparison to~\cite{Antusch:2016con}, we also include a simulation that uses a nearest-neighbor, finite-difference, stencil for the spatial derivatives.
We comment that this method should be less accurate than the pseudospectral method, but we include it so that we can isolate any numerical artifacts that stencil derivatives might have.

%%%%%%%%%%%%%%%%%%%%%%%%%    Results & Discussion  %%%%%%%%%%%%%%%%%%%%%%%%

\begin{figure}[t!]
	\centering
	\includegraphics[width=.99\columnwidth]{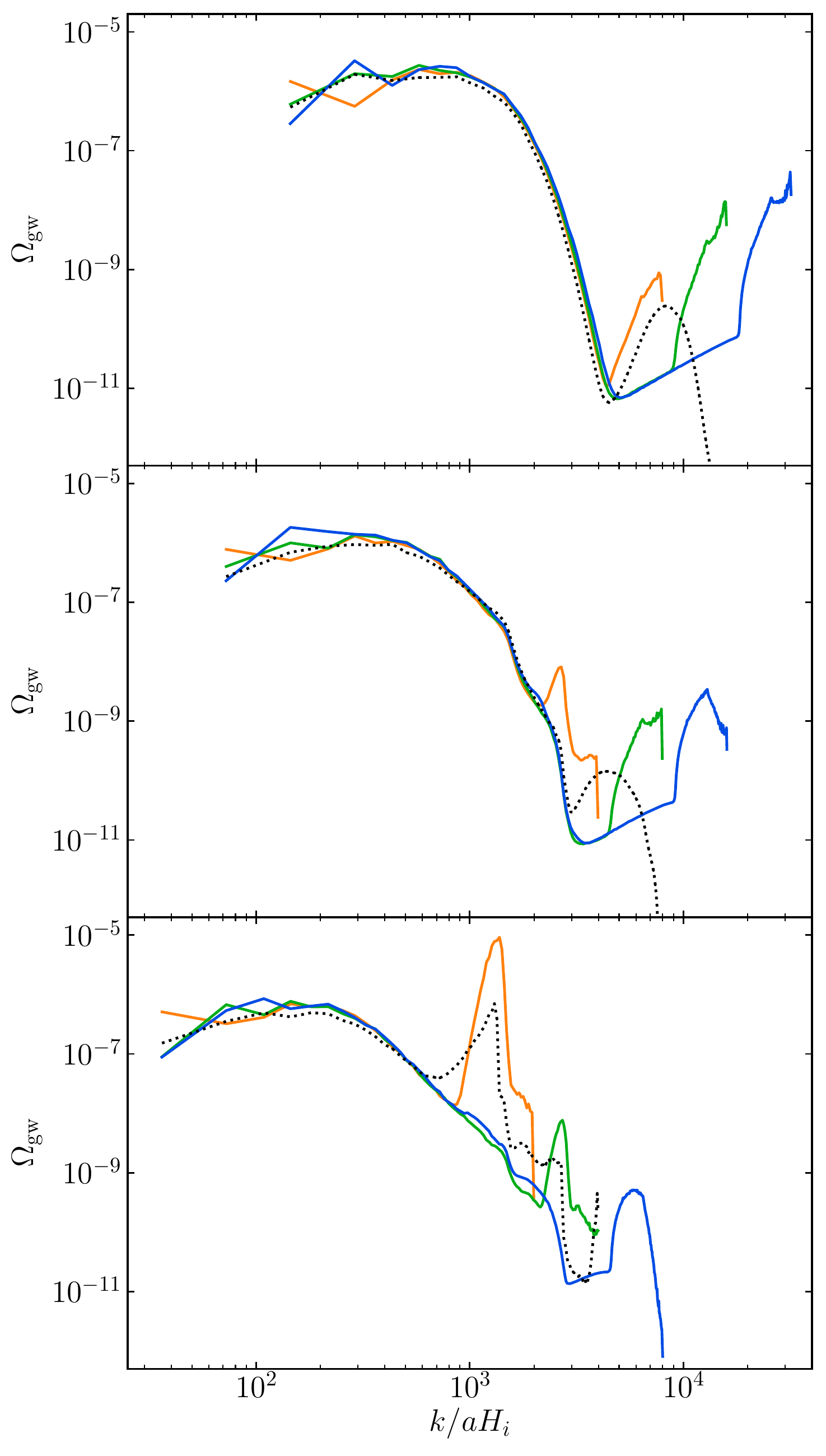}
	\caption{
		Gravitational wave spectra generated over the course of pseudospectral simulations with $N = 64$ (orange), $N = 128$ (green), and $N = 256$ (blue), overlaid with a finite-difference simulation at $N = 128$ (dotted black).
		The first panel corresponds to the point in each simulation when $a(t) = 64/(L_im_{\phi}) = 3.35$, the second panel when $a(t) = 128/(L_im_{\phi}) = 6.7$, and the third when $a(t) \lesssim 256/(L_im_{\phi}) = 13.45$.
		Note that we have not removed the peaks that appear at highest-frequencies in each simulation that come from numerical noise near the Nyquist frequency.
		These peaks correspond to a well-known issue with lattice simulations. Note that the location as well as amplitude of the growing peaks (not just the Nyquist peak) are resolution-dependent, hence not robust physical signatures in the spectrum.
	}\label{fig:spikeplots}
\end{figure}

\begin{figure}[t!]
	\centering
	\includegraphics[width=.99\columnwidth]{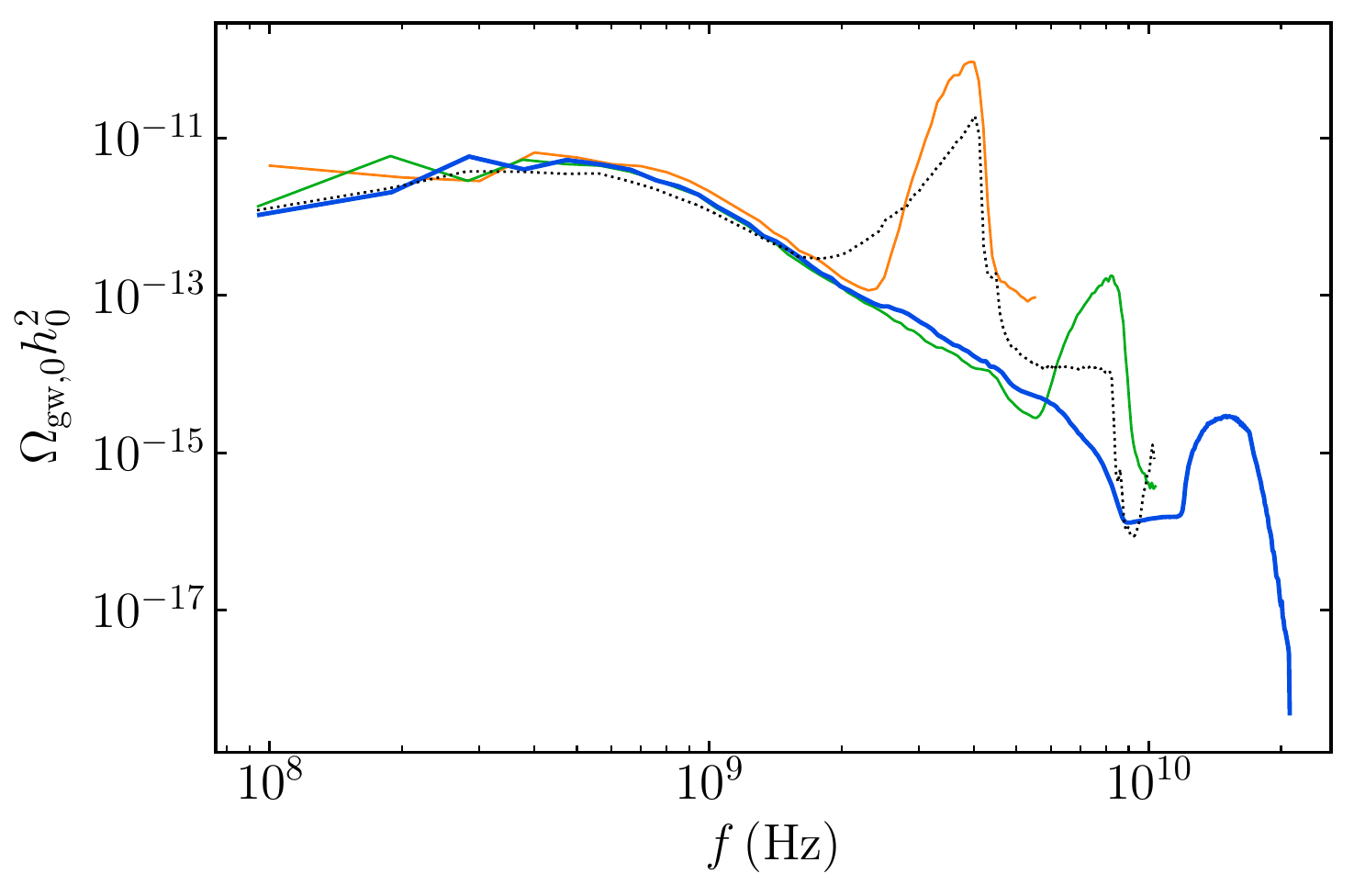}
	\caption{
		Gravitational wave spectra today for $N = 64$ (orange), $N = 128$ (green), and $N = 256$ (blue), overlaid with a finite-difference simulation at $N = 128$ (dotted black).
		We have assumed that gravitational wave production stops at $a=16$ in our simulations, and is followed by radiation domination. We note that apart from the ``shallow" peak at $f\sim 5\times10^8$, the rest of the high frequency peaks are numerical artifacts.}\label{fig:finalSlice}
\end{figure}

\section{Results and discussion}

Starting from vacuum initial conditions for the field perturbations inside our simulation volume, the perturbations in the field grow resonantly.
These perturbations become nonlinear, and the field fragments into oscillons.
The field evolution and fragmentation into oscillons gives rise to gravitational waves.

As with~\cite{Antusch:2016con}, our simulation volume is quite small, and thus may be susceptible to changes in the number of oscillons and their characteristics between different realizations of the initial conditions.
To ameliorate this issue, we performed several simulations at both $N=64$ and $N=128$ using different (random) initial realizations.
Two oscillons form in the majority of our tests, with a few only producing one oscillon.
We did not see significantly different qualitative behavior in any simulation for the same $N$; in particular, the growth of unphysical features in the gravitational wave spectra are nearly identical.

We plot our numerically calculated gravitational wave spectrum $\Omega_{\rm gw}$ in \cref{fig:spikeplots} at the time of production as a function of the physical wave number $k/a$ (in units of $H_i$).
Notice the sharp peak emerging at $k/aH_i \approx 10^3$ in the bottom panel of \cref{fig:spikeplots} for the lowest resolution pseudospectral simulation ($N=64$).
A smaller peak is also seen in the $N=128$ simulation, but now at $k/a H_i\approx 2.5\times 10^3$.
For the highest resolution ($N=256$), a mostly nondynamical peak is seen at $k/aH_i\approx 5\times 10^3$.
We also note that for $N=128$ (the resolution used in~\cite{Antusch:2016con}), the high momentum peak in the finite-difference simulation is clearly at a different momentum and amplitude compared to a pseudospectral simulation at the same resolution.
Hence, none of the high-momentum peaks seen in the bottom panel of \cref{fig:spikeplots} are robust, physical features in the spectra.
In contrast, the spectra at lower momenta are consistent between different resolutions and simulations (the slight differences arise from different realizations of initial perturbations).

While the (mostly) nondynamical peaks related to the Nyquist frequency are already seen in the first panel, the growing peaks begin their growth at different times for different resolutions.
In particular, the growth of the peaks begins when $a(t) dx_{\scriptscriptstyle N} \gtrsim m_\phi^{-1}$ for each $N$ (recall that $dx_{\scriptscriptstyle N}=L_i/N$), strongly suggesting that the results cannot be trusted beyond this point in time.
This is understandable since when $a(t) dx_{\scriptscriptstyle N} \gtrsim m_\phi^{-1}$, unphysical discretization noise pollutes the dynamics on the scales comparable to the size of the objects generating the gravitational wave.

The authors of~\cite{Antusch:2016con} suggest that the peak they see at $k/aH_i\approx 10^3$ would continue to grow past the point they chose to end their simulations; we may extrapolate from our results that in the continuum limit ($N \to \infty$, or $dx \to 0$), this growing peak would never appear.

To be more explicit, we do not find any evidence for a physical, growing peak in the gravitational wave spectrum at $k/aH_i \approx 10^3$ (as claimed in Fig 1. of~\cite{Antusch:2016con}). The growing peak at $k/aH_i\approx 10^3$ that emerges at lower resolutions never appears (or appears at higher physical wave numbers, and/or later times) when we repeat the same simulations at higher resolutions or use different algorithms for numerical evaluation. These results demonstrate that lattice simulations of expanding spacetimes become untrustworthy once the inverse of the physical lattice spacing is larger than the length scales present in a particular problem (in this case the oscillon's spatial extent which is typically $\sim \textrm{few} \times m_{\phi}^{-1}$ in size).

The gravitational wave frequencies in our plots are around $10^9$ Hz (see \cref{fig:finalSlice}), which is mostly set by the energy scale of the universe at the time of production, and, indeed, this is the same for the models of~\cite{Zhou:2013tsa, Antusch:2016con, Antusch:2017flz, Liu:2017hua, Antusch:2017vga}. To lower the frequencies to the LIGO/Virgo bands, it is typically assumed that the scalar field is not the inflaton field but nevertheless develops a VEV during the early stages of its evolution. This field eventually dominates the universe and then begins to fragment into oscillons in the relatively late stages of the early universe.

In earlier studies of gravitational wave production from oscillons in symmetric potentials~\cite{Zhou:2013tsa}, a {\it growing peak} like the one suggested by~\cite{Antusch:2016con} was not observed. A peak is observed early in the simulations of~\cite{Zhou:2013tsa} from the formation of oscillons, but stops growing at late times.
The authors of~\cite{Antusch:2016con, Antusch:2017vga} argue that this is because asymmetric potentials lead to highly non-spherical oscillons which source gravitational waves for a long time after formation. A higher resolution simulation carried out by the same authors in~\cite{Antusch:2017vga} shows a much weakened peak for the same models, which the authors attribute to potential differences in initial conditions/numerical artifacts.

We do not completely understand the connection between asymmetry of the potential in field space and the asymmetry of the oscillons in physical space. Nevertheless, even if there is a connection, we see no evidence of a {\it physical}, growing peak in gravitational waves from the dynamics of such asymmetric oscillons in our simulations for the models and parameters used in~\cite{Antusch:2016con}. Moreover, aspherical oscillons should become more spherical through emission of scalar radiation. We note the rate of sphericalization, albeit in a different model and in 2D, was studied in~\cite{Hindmarsh:2006ur}, which is shorter than the timescale of the simulations.

High frequency gravitational wave generation from the production and dynamics of localized, nonexpanding sources such as oscillons provide an exciting possibility for probing physics of the early universe. Features in the gravitational wave spectrum can exist that indicate the presence of oscillons~\cite{Zhou:2013tsa,Antusch:2017vga,Liu:2017hua}. However, as we have seen, special care is needed in interpreting results from numerical simulations whose resolution inevitably reaches the physical size of the sources in an expanding universe. We will address these issues for a network of oscillonlike sources in future work.

%%%%%%%%%%%%%%%%%%%%%%%%%%%%%%%%%%%%%%%%%%%%%%%%%%%%%%%%%%%

\acknowledgments

We thank S. Antusch, F. Cefal\`a, S. Krippendorf, K. Lozanov, F. Muia, and S. Orani for helpful discussions.
M.A. is supported by a United States Department of Energy Grant No. DE-SC0018216.
J.B. is supported by the European Research Council under the European Community's Seventh Framework Program (FP7/2007-2013) / ERC grant agreement no 306478-CosmicDawn.
E.J.C. acknowledges financial support from STFC consolidated grant No. ST/L000393/1 and ST/P000703/1.
J.T.G. and C.S. are supported by the National Science Foundation Grant No. PHY-1719652.
Z.J.W. is supported in part by the United States Department of Energy Computational Science Graduate Fellowship, provided under Grant No. DE-FG02-97ER25308. SYZ acknowledges support from the starting grant from University of Science and Technology of China (KY2030000089) and the National 1000 Young Talents Program of China.
Part of this work was performed at the Aspen Center for Physics, which is supported by National Science Foundation grant PHY-1607761.
Numerical simulations were performed on equipment provided by the Kenyon Department of Physics, the National Science Foundation and NASA under NASA Astrophysics Theory Grant No. NNX17AG48G.

%%%%%%%%%%%%%%%%%%%%%%%%%%%%%%%%%%%%%%%%%%%%%%%%%%%%%%%%%%%

\bibliographystyle{apsrev4-1} % Tell bibtex which bibliography style to use
\bibliography{oscillon-gw}

\end{document}